\documentclass[
]{ceurart}

\usepackage{listings}
\usepackage{multirow}
\usepackage{array}
\usepackage{graphicx}
\usepackage{pdfpages}
\begin{document}

\copyrightyear{2025}
\copyrightclause{Copyright © 2025 for this paper by its authors. Use permitted under Creative Commons License Attribution 4.0 International (CC BY 4.0).}

\conference{Joint Proceedings of the ACM IUI Workshops 2025, March 24-27, 2025, Cagliari, Italy}

\title{Explainable Biomedical Claim Verification with Large Language Models }

\author[1]{Siting Liang}[%
orcid=0009-0000-0146-581X,
email=siting.liang@dfki.de,
]
\address[1]{German Research Center for Artificial Intelligence, Germany}

\author[1,2]{Daniel Sonntag}[%
orcid=0000-0002-8857-8709,
email=daniel.sonntag@dfki.de,
]
\address[2]{University of Oldenburg, Germany}


\begin{abstract}
  Verification of biomedical claims is critical for healthcare decision-making, public health policy and scientific research.  We present an interactive biomedical claim verification system by integrating LLMs, transparent model explanations, and user-guided justification. In the system, users first retrieve relevant scientific studies from a persistent medical literature corpus and explore how different LLMs perform natural language inference (NLI) within task-adaptive reasoning framework to classify each study as "Support," "Contradict," or "Not Enough Information" regarding the claim. Users can examine the model's reasoning process with additional insights provided by SHAP values that highlight word-level contributions to the final result. This combination enables a more transparent and interpretable evaluation of the model's decision-making process. A summary stage allows users to consolidate the results by selecting a result with narrative justification generated by LLMs. As a result, a consensus-based final decision is summarized for each retrieved study, aiming safe and accountable AI-assisted decision-making in biomedical contexts. We aim to integrate this explainable verification system as a component within a broader evidence synthesis framework to support human-AI collaboration.
\end{abstract}


\maketitle

\section{Introduction}
Automated biomedical claim verification systems aim to assist clinicians and researchers in combating the potential harm of misinformation in the healthcare domain. These systems verify claims related to treatments, clinical trial outcomes, and other medical assertions by synthesizing evidence from clinical trial data and scientific literature. Biomedical claim verification involves assessing the veracity of such claims through relevant studies, ensuring reliable conclusions for critical decision-making \cite{wadden-etal-2020-fact, sarrouti-etal-2021-evidence-based, saakyan2021covid, jullien-etal-2023-semeval}. Research in biomedical claim verification has focused on utilizing advanced natural language processing (NLP) techniques. Fine-tuned natural language inference (NLI) models have been widely adopted for this task \cite{guo2022survey}. These models typically follow a standard pipeline: (1) retrieve relevant studies using the claim as query, (2) process the claim and retrieved evidence using a language model in either a fine-tuned or in-context learning setup designed for NLI task \cite{wadden2021multivers, liu2024retrieval}. The NLI task requires determining the logical relationship between two pieces of text: a premise (in our case, scientific studies) and the biomedical claim to be verified. The task involves classifying this relationship into one of three categories: \textsc{Support}, \textsc{Contradict} and \textsc{No Enough Information}. Related examples from \citet{wadden-etal-2020-fact} are shown in Table~\ref{table:claims-relations}. In the scientific and medical domains, NLI models are required to process long and complex documents while also a deep understanding of biomedical knowledge to interpret specific terminologies \cite{romanov2018lessons, wadden2021multivers, liu2024retrieval}. In particular, scientific studies often contain complex statistical information and precise measurements that must be interpreted accurately to avoid errors in claim verification \cite{jullien-etal-2024-semeval}. Large language models (LLMs) offer promising potential to address these challenges \cite{jullien2024semeval}. Their effectiveness depends on two critical factors: the size of the model and the suitability of the prompts designed for specific tasks \cite{huang2022towards, qiao2022reasoning, xia2024chainofthoughtsurveychainofxparadigms}.

\begin{table}[h]
\centering
\begin{tabular}{|p{3cm}|p{8cm}|p{2cm}|}
\hline
\textbf{Claim}       & \textbf{Most Relevant Study}                         & \textbf{Relation} \\
\hline
\textcolor{orange}{76-85\% of people} with severe mental disorder receive no treatment in low and middle income countries. &
\multirow{2}{8cm}{... RESULTS The prevalence of having any WMH-CIDI/DSM-IV disorder in the prior year varied widely, from 4.3\% in Shanghai to 26.4\% in the United States.... \textcolor{orange}{Although disorder severity was correlated with probability of treatment in almost all countries, 35.5\% to 50.3\% of serious cases in developed countries and 76.3\% to 85.4\% in less-developed countries received no treatment in the 12 months before the interview.} Due to the high prevalence of mild and subthreshold cases, the number of those who received treatment far exceeds the number of untreated serious cases in every country.} &
Support / \newline Entailment\\
\cline{1-1} \cline{3-3}
\textcolor{orange}{10-20\% of people} with severe mental disorder receive no treatment in low and middle income countries. &  & Contradict \\
\hline
\end{tabular}
\caption{Examples of biomedical claim verification illustrating the logical relationship between two claims and the retrieved most relevant study. Phrases highlighted are the critical statistic information that determines the logical relation between the claims and study.}
\label{table:claims-relations}
\end{table}

Real-world applications of LLMs, especially in healthcare and scientific domains, demand high levels of transparency, interpretability, and trustworthiness due to the high-stakes nature of decisions \cite{miller2019explanation, kotonya2020explainable, huang2024trustllm}. In this work, we propose an interactive biomedical claim verification system as part of the \textit{No-IDLE} \cite{pub15026} and \textit{No-IDLE meet ChatGPT} \cite{pub15382} projects about interactive deep learning and LLMs. 
The primary functionality of the system is to assist users in validating claims by leveraging the strengths of LLM-based verification while ensuring a transparent and reliable decision-making process. The system builds on the Chain of Evidential Natural Language Inference (\textbf{CoENLI}) framework, which enables LLMs to generate evidence-based rationales before arriving at a final relation classification. To evaluate the framework, we use two relevant biomedical benchmarks, demonstrating that (\textbf{CoENLI}) significantly improves LLM accuracy and outperforms the general Chain of Thought (CoT) approach \cite{wei2022chain}. The evaluation results demonstrate that by explicitly outlining the evidence-based reasoning process, \textbf{CoENLI} enhances both the interpretability and reliability of claim verification. 

To further enhance the interpretability of the system, we integrate \textsc{Shapley Additive exPlanations} (SHAP) saliency maps \cite{lundberg2017unified}, which highlights the word-level contributions within the generated rationales. This technique provides a deeper understanding of how LLMs weigh specific evidence for arriving at final conclusion. In addition, the system employs different LLMs to provide users with a comparative analysis of results and reasoning. By enabling users to review different perspectives and reasoning outputs, the system fosters a nuanced understanding of the claim verification process, ultimately increasing trustworthiness of LLMs in real-world applications. Finally, users select the most appropriate classification as the final decision after reviewing model-generated rationales and explanations. Our main contribution in this work is the development of an iterative human-AI collaboration workflow that ensures transparency, accountability, and adaptability to individual expert knowledge. By leveraging LLMs with \textbf{CoENLI}, the system delivers transparent evidence-based evaluations while incorporating SHAP saliency explanations. Additionally, comparative insights from multiple LLMs further enhance understandability and reliability. Together, these innovations advance automatic biomedical claim verification towards greater confidence and usability in the process.

\section{Explainable Biomedical Claim Verification System}
\subsection{Overview}
\begin{figure}[htp] 
\centering
\includegraphics[angle=90,width=\textwidth,, height=\textheight]{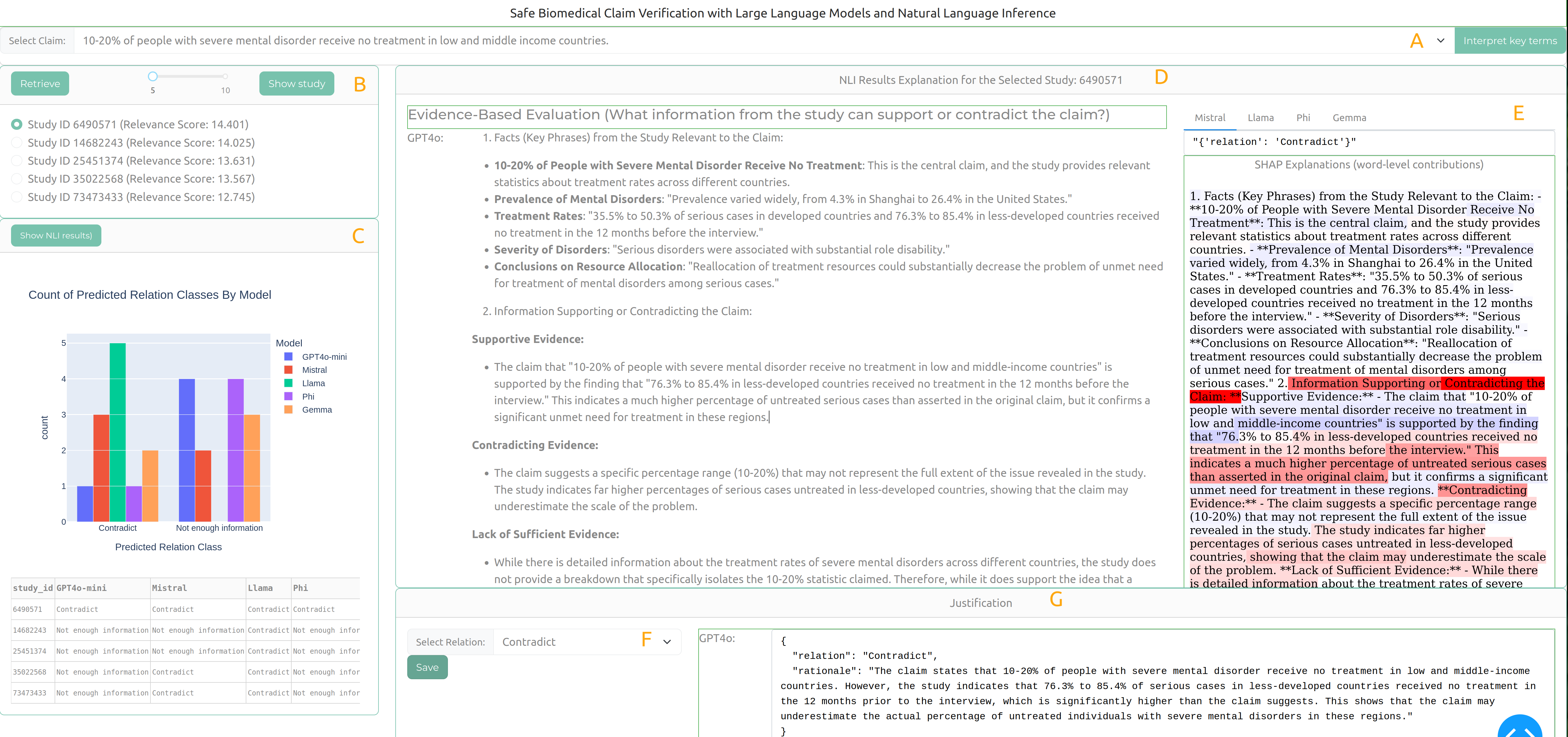}
\caption{The biomedical claim verification system comprises several interactive components for the user study. }
\label{fig:system}
\end{figure}

The system comprises several interactive components to combine the strengths of advanced language models-driven analysis and user control. Figure~\ref{fig:system} provides an overview of our system's user interface. Users initiate the verification process by selecting a claim to investigate (A) and retrieving relevant studies from database of scientific literatures using BM25 algorithm \cite{robertson2009probabilistic} to the claim of being assessed (B). The system employs multiple LLMs within the \textbf{CoENLI} framework to evaluate the relationship between the claim and each retrieved study (C).

\begin{figure}[h!]
\centering
\includegraphics[width=\textwidth]{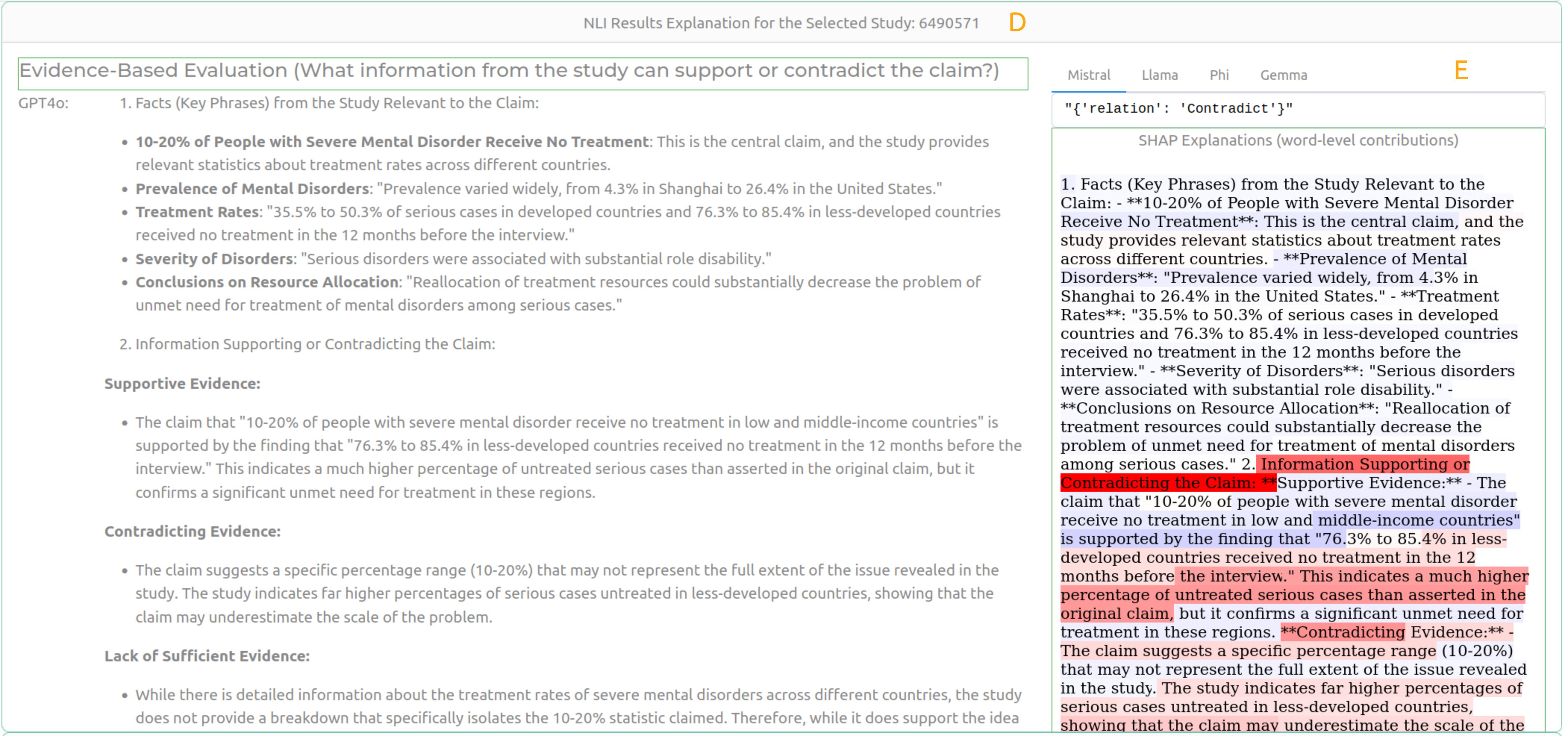}
 \caption{Components D and E provide users dual-layer interpretability for a deeper understanding of verification results by combining evidence analysis with SHAP-based rationale highlighting.}\label{fig:d}
\end{figure}

\newpage
 To enhance the transparency of the verification results, the interface displays model's analysis of evidence from the selected study that supports or contradicts the claim (D). To enhance understanding, we use SHAP values to highlight which parts of the rationales contributed most to the model’s final decision, revealing the model’s focus in drawing the conclusion (E) . This dual-layer interpretability, showing both how the model analyzes evidence and how it arrives at its final classification, provides users with a deeper understanding of the verification results (see Figure~\ref{fig:d}). 

\begin{figure}[h!]
\includegraphics[width=\textwidth]{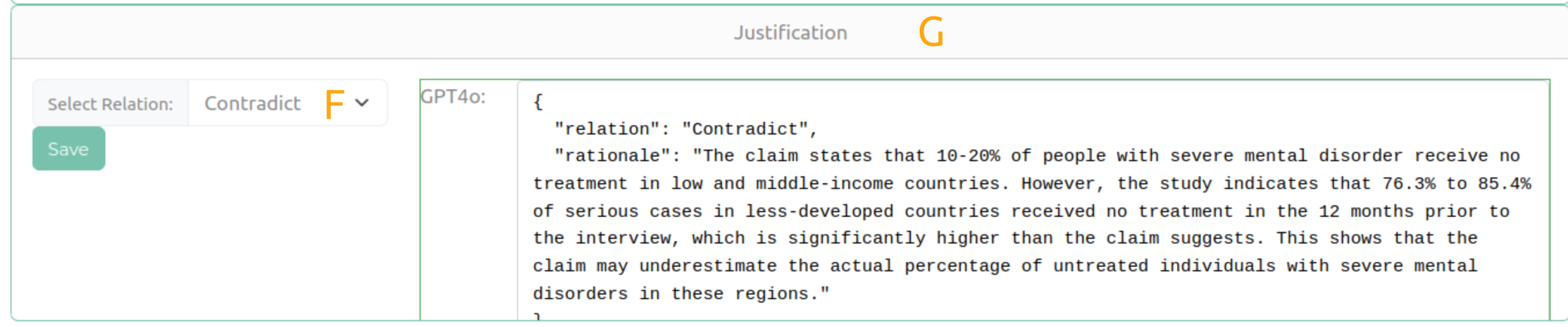}
 \caption{Components F and G comprise the summary stage, where users actively engage by adjusting the classification results and prompting the model to generate a final justification for the final decision.}\label{fig:f}
\end{figure}

After reviewing the generated rationales and saliency maps, if the users disagree with the initial classification result, they can adjust it (F), prompting the model to generate a concise justification for the updated classification (G) (see Figure~\ref{fig:f}). 

This iterative approach allows for user involvement and enhanced trustworthiness in claim verification. More details about the \textbf{CoENLI} framework and SHAP values are explained in the subsections~\ref{sec:coenli} and ~\ref{sec:shap}. The evaluation of the \textbf{CoENLI} framework and the choice of models are to be found in the section \ref{sec:eval}.  Our study of the explainable claim verification system aims to assist human experts in making informed decisions by clearly presenting the evidence analyzed by LLMs while also enhancing transparency and providing comprehensive model explanations. The ultimate goal is to enable users to effectively assess the rationale behind the system's conclusions, fostering trust and facilitating collaboration in complex decision-making tasks. 

\subsection{Chain of Evidence-Based Natural Language Inference (CoENLI)}
\label{sec:coenli}

\begin{figure}[h!]
\includegraphics[width=\textwidth]{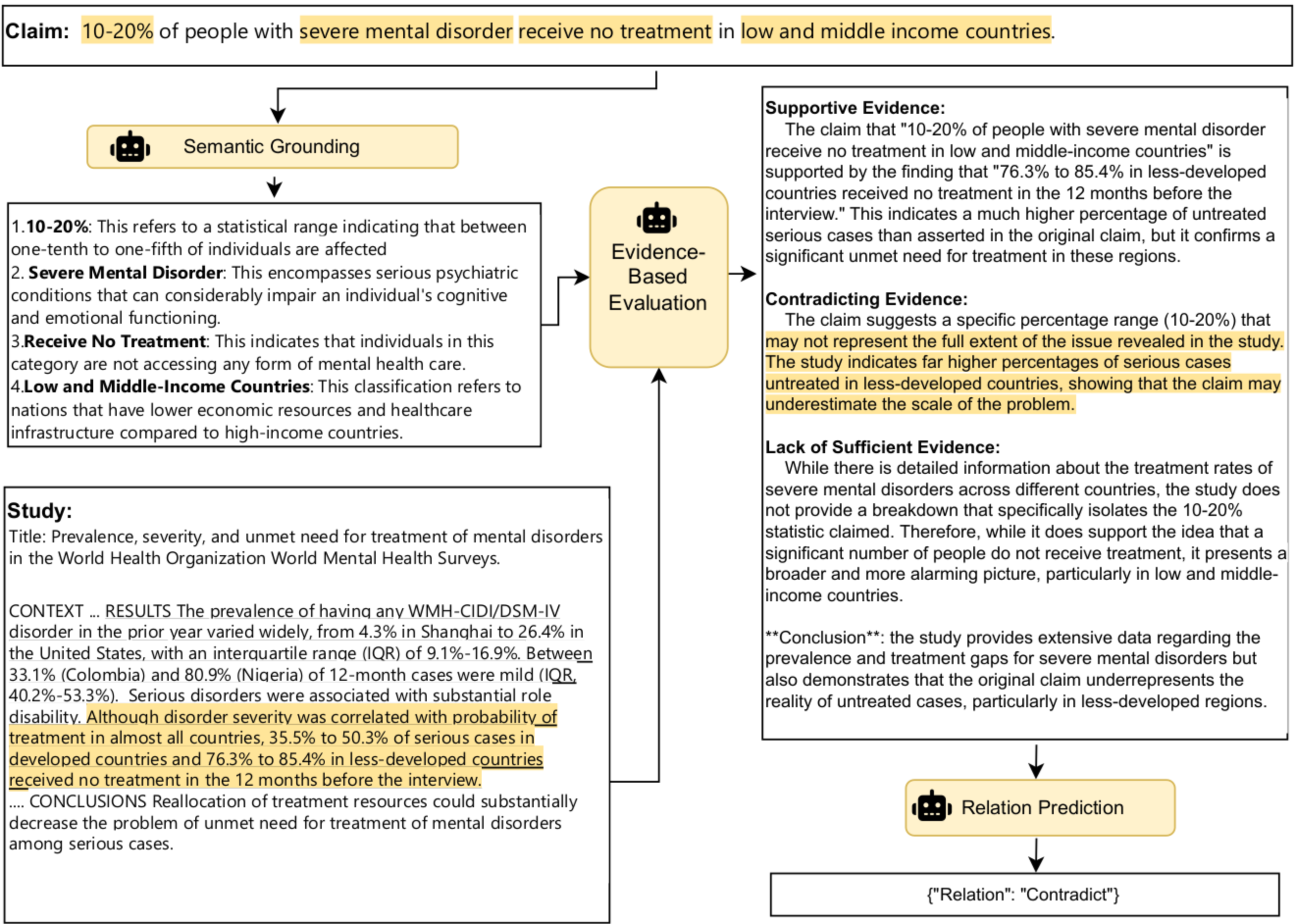}
\caption{When prompting the LLMs with \textbf{CoENLI} framework, the process begins with \textit{Semantic Grounding} and \textit{Evidence-based Evaluation} steps. These steps help interpret key terms and assess each piece of claim against identified relevant data points. The highlighted words and phrases in the claim, study, and generated evaluation are intended to offer plausible insights involved in the claim verification process.}
\label{fig:coenli}
\end{figure}

When prompting LLMs to complete reasoning tasks, breaking down complex reasoning tasks into simpler steps can be useful. Chain-of-Thought (CoT) strategies \cite{wei2022chain, kojima2022large}, which provide exemplars of reasoning processes have demonstrated impressive performance across different reasoning benchmarks. Decomposition steps are useful for increasing the reliability of model generations \cite{yu2023towards}.  \citet{zhou2022least} noted that step-wise prompting require task-specific design for optimal performance. Inspired by prior works  \cite{zhou2022least, lei2023chain}, we propose \textbf{CoENLI} to refine the CoT reasoning in claim verification tasks including the following steps and Figure~\ref{fig:coenli} illustrates the reasoning process within \textbf{CoENLI} with the example from Table~\ref{table:claims-relations}. 
\begin{itemize}
\item Semantic Grounding: A task instruction prompt contains the phrase \textit{"Interpret the key terms in the claim"}. It activates specific semantic understanding of biomedical knowledge and logical patterns in LLMs, providing a contextual foundation for the subsequent reasoning step.

\item Evidence-Based Evaluation: In this step, the model extracts relevant evidence from the premise data (e.g., scientific studies) and systematically evaluates the claim by comparing its key elements with the extracted evidence. The process is guided by instruction prompts such as: \textit{"1. Identify the relevant facts in the study. 2. Evaluate each piece of information in the claim against the facts."}.
\item Relation Prediction: In the final step, the reasoning process concludes with a concise classification (e.g., \textsc{Support} or \textsc{Contradict}) expressed in natural language. This prediction is based on the previously generated terms interpretations and evidence analysis, which are sequentially chained into the input prompt to guide the model’s final decision.
\end{itemize}

\subsection{SHAP Values for Interpreting Word-Level Contributions} 
\label{sec:shap}
\begin{figure}[h!]
\includegraphics[width=\textwidth]{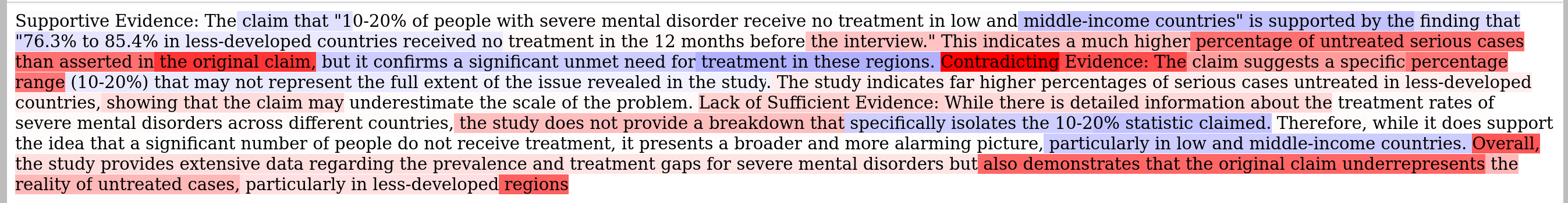}
\centering
\caption{Words with positive SHAP values (highlighted in red) indicate phrases within the model-generated evaluation that significantly contribute to the \textsc{Contradict} classification, while words with negative SHAP values (highlighted in blue) indicate elements that reduce the likelihood of this classification. }
\label{fig:shap}
\end{figure}

\textbf{CoENLI} enables LLMs to generate intermediate, evidence-based rationales and provide human-readable explanations of how the model processes claims and evidence. As illustrated in Figure~\ref{fig:coenli}, the \textit{Evidence-Based Evaluation} step allows for broad reasoning, including identifying relevant evidence and evaluating both supportive and contradictory information. However, consolidating this evaluation into a final relation remains opaque, leaving interpretability gaps about which aspects of the evaluation contribute most to the final result. To address these limitations, we incorporate SHAP explanations using modules from \cite{shap}\footnote{\url{https://shap.readthedocs.io/en/latest/text_examples.html}}, specifically developed to explain language models. By analyzing the Shapley values (SHAP) associated with the words in the input prompt, we can identify which features (words or phrases) generated in the intermediate step had the most significant influence on the final output of the generative model. Figure~\ref{fig:shap} illustrates feature relevancy based on SHAP values uisng a Mistral model \cite{jiang2023mistral} as explainer for the final relation result of the example from Figure~\ref{fig:coenli}. The saliency maps provide detailed insights into the model's decision-making process when balancing supportive and contradictory evidence to determine the relationship between the example claim and the study. 

\section{Evaluation}
\label{sec:eval}
 Biomedical claim verification can be defined as logical relationship classification problem, where an NLI model determines whether a claim (\(C\)) logically follows from the premise evidence (\(P\)) provided in clinical trial or scientific study data. In our evaluation, we denote:
\begin{equation}
\small
f(C, P) = 
\begin{cases} 
\text{Support} & \text{if \( C \) logically follows} \\ 
                                        & \text{ from \( P \);} \\
\text{Contradict} & \text{otherwise}
\end{cases}
\end{equation}

\subsection{Datasets and Models}
 In order to assess the generalization capabilities of \textbf{CoENLI} across different LLMs, we evaluate their performance on two related benchmarks  \textbf{NLI4CT} \cite{jullien-etal-2023-semeval, jullien-etal-2024-semeval} and \textbf{SciFact} \cite{wadden-etal-2020-fact, wadden-etal-2022-multivers}. The claims from both datasets are written by human experts given clinical trial reports or scientific studies respectively. The \textbf{NLI4CT} challenges highlighted the difficulties of applying NLI models to validate claims related to clinical trial reports (CTRs). It requires a deep understanding of medical and scientific knowledge to interpret implicit data points beyond simple text matching. The premises in \textbf{SciFact} consist of evidence sentences extracted from the abstract of relevant studies. \citet{wadden-etal-2022-multivers} demonstrated the advantages of incorporating document-level premises compared to  sentence-level premises for the \textbf{SciFact} challenge. Table~\ref{tab:dataset_statistics} summarizes the number of instances for each relation class of the datasets applied in our evaluation.
 
\begin{table}[h!]
\centering
\scalebox{0.90}{
\begin{tabular}{|l|c|c|c|}
\hline
\textbf{Dataset} & \textbf{Support / Entailment}  & \textbf{Contradict} \\ \hline
\textbf{NLI4CT} (test set) & 250 & 250  \\ \hline
\textbf{SciFact} (dev set) & 216  & 122  \\ \hline
\end{tabular}}
\caption{The number of evaluation instances. \textbf{SciFact}’s test set withholds ground truth labels for leaderboard submissions \cite{wadden-etal-2020-fact}, here we use its dev set as substitute.}
\label{tab:dataset_statistics}
\end{table}

In order to increase the trustworthness of LLMs' outcomes, our system employs different lightweight open-source LLMs \cite{abdin2024phi,jiang2023mistral,team2024gemma, dubey2024llama}. The models applied are instruction-tuned \cite{ouyang2022training} and compatible with the \textit{FastLanguageModel} modules of unsloth.ai \cite{unsloth} for faster running on a single NVIDIA A100–80GB GPU. Table~\ref{tab:models} in Appendix~\ref{app:models} provides the version information about the applied models. In the evaluation, we compared their performance with two low-cost GPT models \cite{Models}. 


\subsection{Results}
To evaluate the reasoning capabilities of LLMs in a straightforward manner, We report prediction accuracy using F1 scores as detailed in Table~\ref{tab:zero_shot_results}. 
\[
F1 = 2 \cdot \frac{\text{Precision} \cdot \text{Recall}}{\text{Precision} + \text{Recall}}
\]
The F1 scores are calculated using the scikit-learn library\footnote{\url{https://scikit-learn.org/stable/modules/generated/sklearn.metrics.f1_score.html}}. Our evaluation compares the performance of \textbf{CoENLI} against two baseline prompting methods:
\begin{itemize}
    \item Simple prompt: A task-specific prompt template
\textit{"Return the logical relation between the provided claim and study: <Support> or <Contradict>."}. This represents a minimal and direct approach to the task with LLMs.

\item zero-shot CoT: Building on the simple prompt, we introduce an additional instruction: \textit{"Evaluate the relation step by step."} as proposed by \citet{kojima2022large}, prompting LLMs to deliver an intermediate reasoning process before responding the final relationship.
\end{itemize}

The comparison highlights the impact of task-adaptive \textbf{CoENLI} framework on the prediction accuracy of LLMs.

\begin{table}[h!]
\centering
\scalebox{0.90}{
\begin{tabular}{lcccccccc}
\toprule
\textbf{Model} & \multicolumn{4}{c}{\textbf{NLI4CT}} & \multicolumn{4}{c}{\textbf{SciFact}} \\
\cmidrule(lr){2-5} \cmidrule(lr){6-9} 
& Simple & CoT & \textbf{CoENLI} &\textbf{CoENLI$^*$}  & Simple & CoT & \textbf{CoENLI}   &\textbf{CoENLI$^*$}\\
\midrule
\textbf{GPT3.5} & 0.52 ± 0.01 & 0.53 ± 0.00 & 0.75 ± 0.01 & 0.82 ± 0.00 & 0.51 ± 0.03 & 0.76 ± 0.00 & 0.86 ± 0.00 &  0.88 ± 0.01 \\
\textbf{GPT4o-mini} & 0.67 ± 0.01 & 0.77 ± 0.02 & 0.86 ± 0.01 & - & 0.83 ± 0.01 & 0.88 ± 0.00 & 0.94 ± 0.01 & - \\
\midrule

\textbf{Llama3.1-8B}& 0.47 ± 0.00 & 0.54 ± 0.01 & 0.67 ± 0.02 & 0.80 ± 0.00 & 0.53 ± 0.02 & 0.80 ± 0.01 & 0.84 ± 0.05 &  0.90 ± 0.01 \\
\textbf{Gemma2-9B}& 0.63 ± 0.00 & 0.67 ± 0.03 & 0.75 ± 0.03 &  0.80 ± 0.00& 0.57 ± 0.01 & 0.73 ± 0.00 & 0.86 ± 0.02  & 0.89 ± 0.01 \\
\textbf{Mistral-12B}& 0.55 ± 0.00 & 0.65 ± 0.01 & 0.75 ± 0.01 &  0.82 ± 0.00 & 0.65 ± 0.01 & 0.83 ± 0.00 & 0.87 ± 0.02 & 0.89 ± 0.00  \\
\textbf{Phi3-14B}& 0.62 ± 0.01 & 0.64 ± 0.00 & 0.75 ± 0.02 &  0.82 ± 0.00 & 0.76 ± 0.03 & 0.80 ± 0.01 & 0.88 ± 0.02 & 0.90 ± 0.01 \\
\bottomrule
\end{tabular}}
\caption{F1 Scores (mean ± standard deviation) for three benchmarks in zero-shot scenario. We compare the performance across the cost-effective GPT models and open sourced lightweight LLMs. \textbf{CoENLI$^*$} represents the results of applying GPT4o-mini in the \textit{Evidence-Based Evaluation} step. All the results demonstrate the high accuracy of the inference capability of GPT4o-mini model in the \textbf{CoENLI} framework in zero-shot setting.}
\label{tab:zero_shot_results}
\end{table}

While \textbf{CoENLI} demonstrates the enhanced performance of LLMs in the claim verification task compared to the baseline methods, a performance gap persists between GPT4o-mini and small-scale LLMs. Lightweight LLMs face challenges in delivering high-quality text analysis and often require fine-tuning with additional task-specific training examples \cite{ding-etal-2023-enhancing}. Despite these limitations, smaller open-source models offer flexibility for SHAP explanations and can be optimized for specific tasks through the collection of training data during long-term development, increasing the controllability of the system. The choice between fine-tuning lightweight LLMs or incorporating GPT4o-mini’s evaluations into the \textbf{CoENLI} pipeline depends on the application’s requirements and resource constraints. For our user study, we leverage GPT4o-mini's robust reasoning capabilities for the \textit{Evidence-Based Evaluation} step. These outputs are then passed to small-scale models, which focus on generating final decisions and adding a second layer of interpretability using SHAP values. This hybrid approach combines the strengths of advanced and lightweight models to achieve both accuracy and transparency. 

\subsection{User Study}
In constructing the datasets, \citet{wadden-etal-2020-fact} reported a Cohen's kappa of 0.75 as inter-annotator agreement. Similarly, \citet{jullien-etal-2023-semeval} conducted a human evaluation of the \textbf{NLI4CT} task with three experts, achieving an average accuracy of 85\% against the gold labels, the inter-annotator agreement with a Cohen's kappa of 0.83 . These results highlight the inherent variability in human judgement. This variability is often due to task uncertainty and differences in individual knowledge.

To evaluate the utility of the explainable claim verification system, we employ four medical students, each tasked with  selecting a verdict for 20 claims from the \textbf{SciFact} test set (which lacks ground truth annotations) against five retrieved studies independently, resulting in 100 claim-study pairs in total. The user interface used for the study is shown in Figure \ref{fig:system}. We observed an increase in inter-annotator agreement, with Cohen’s kappa rising from 0.74 (without the Evidence-Based Explanation and SHAP values explanation) to 0.81. This improvement suggests that the system’s enhanced interpretability fosters trust in the LLM outcomes and less modification of the initial results generated by the LLM, therefore promoting better alignment among users.

To assess the transparency and understandability of the model's reasoning process leading to the final relation classification, participants rated the model-generated rationales on a scale from 1 to 5. A score of 1 indicated that the reasoning process was confusing, with no clear connection between the rationales and the final classification, while a score of 5 indicated that the reasoning was fully transparent and easy to follow. All participants rated the reasoning transparency as a \textbf{4}, suggesting that the model's reasoning process is generally perceived as transparent. When asked what could be improved, participants provided feedback indicating the need for better quality in the intermediate reasoning steps generated by the LLMs. One participant noted, \textit{"The NLI model sometimes overlooked the smallest details in the claim."}. This highlights how enhanced interpretability can help identify limitations in reasoning of LLM. As emphasized by \cite{huang2024position}, improving the functionality of these model-generated explanations is crucial for fostering user confidence in the system.

\subsection{Feedback Loop Integration}
\begin{figure}[htp]
\includegraphics[width=0.8\textwidth]{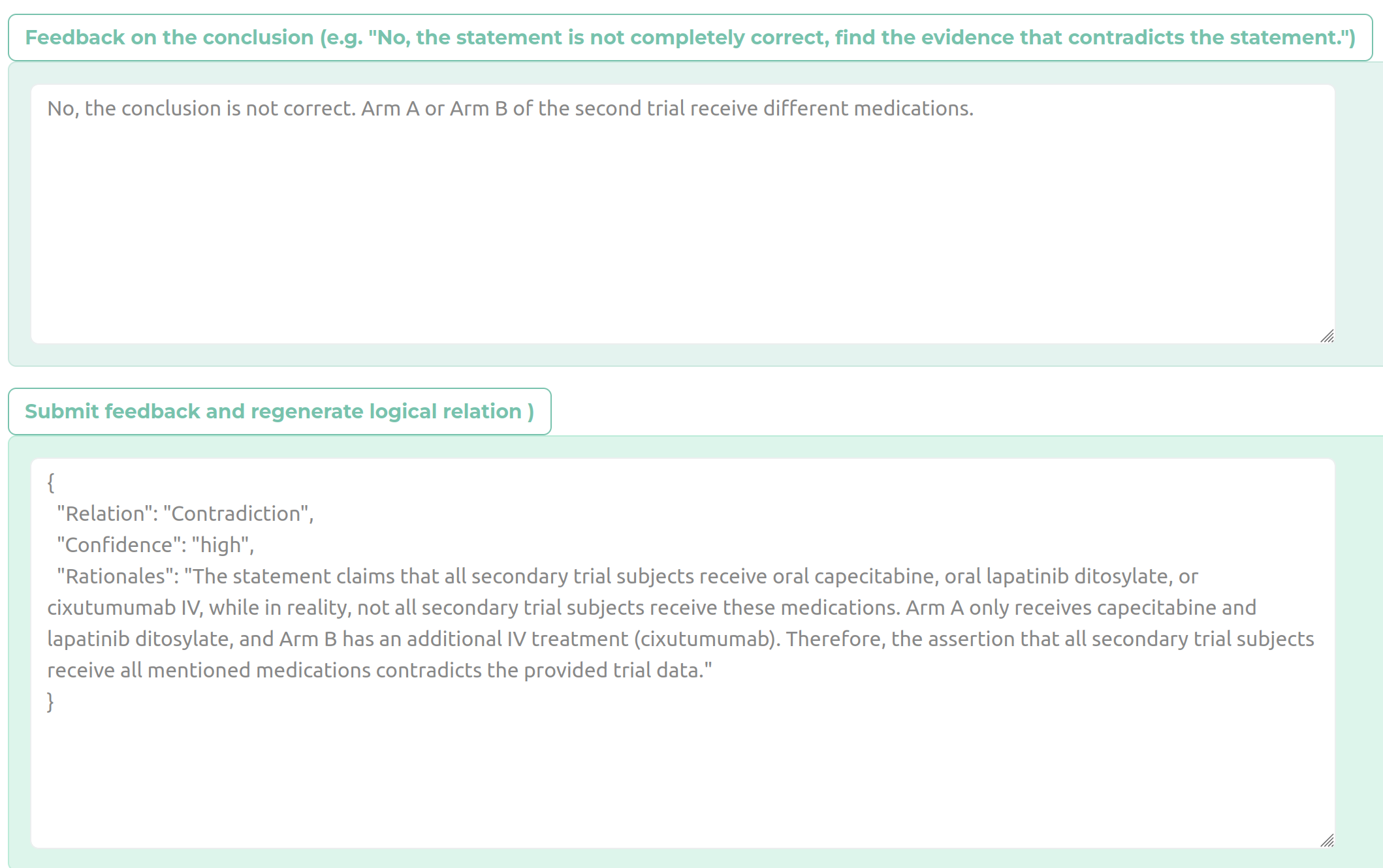}
\centering
\caption{Consider the following verification result of one claim - \textit{"All the primary trial participants do not receive any oral capecitabine, oral lapatinib ditosylate or cixutumumab IV, in conrast all the secondary trial subjects receive these."} from \textbf{NLI4CT}: \textit{\{"relation": "Entailment", "rationales": "The claim can be logically inferred from the clinical trial data points provided. The data clearly show that primary trial participants do not receive any oral capecitabine, oral lapatinib ditosylate, or cixutumumab IV, while all secondary trial subjects receive these medications in either Arm A or Arm B of the trial. This conclusion is supported by the specific interventions outlined for each group in the trial."\}}. However, this result overlooks quantitative consistency—specifically the mismatch between "all the secondary trial subjects" in the claim and the different medications in "Arm A" and  "Arm B" of the secondary trial subjects. This inconsistency highlights the difficulty of accurately aligning numerical or quantifier details within the reasoning process of LLMs. Hence, we integrate a feedback loop to empower users to actively refine the model’s reasoning process.}
\label{fig:feedback}
\end{figure}

Based on the results of the user study, which revealed that the model sometimes misses small details in claim verification, we integrate a feedback loop into the verification system. This enhancement directly addresses the identified gaps by allowing users to provide detailed feedback on specific errors or overlooked aspects of the model's reasoning. Through this mechanism as illustrated in Figure \ref{fig:feedback}, users can highlight inconsistencies or missing details with concise guidance, enabling the system to regenerate both the relationship classification and the corresponding justifications. This iterative process not only helps to correct errors in real time, but also guides the model to refine its reasoning, ultimately improving its accuracy and reliability in verifying biomedical claims.  

Furthermore, these feedback-driven rationales can be leveraged to fine-tune the model, enhancing its ability to recognize and account for nuanced details, particularly in logical consistency comparisons. This human-in-the-loop approach fosters greater user engagement, promotes collaborative verification, and ultimately strengthens the reliability of LLM-based biomedical claim verification systems.

\newpage 
\section{Related Work}
Biomedical claim verification falls into the broader task of \textsc{Fact-Checking} \cite{vlachos-riedel-2014-fact}. Automated claim verification is seen as a potential solution to enhance the speed and comprehensiveness of fact-checking in high-demand healthcare field \cite{kotonya2020explainable, wang2023check}. Additional datasets have been constructed and advanced machine learning methods to drive progress in automated biomedical claim verification system \cite{wadden-etal-2020-fact, saakyan2021covid, sarrouti-etal-2021-evidence-based, jullien-etal-2023-semeval}. However, in real-world scenarios, the verdict of claims is rarely either \textsc{Support} or \textsc{Contradict}, but often partially correct, contextually dependent, or misleading without additional explanation. \citet{nakov2021automated} argued that automated claim verification systems may aim to provide nuanced understanding rather than binary classifications. \citet{li2023self} introduced a self-checker framework leveraging LLMs, which includes an evidence-seeker module to extract relevant evidence sentences for a given claim from retrieved passages. The framework allows human workers to validate the verdict prediction alongside the presented evidence, ensuring a more reliable verification process.

\citet{chen2024combating} discussed the opportunities and challenges introduced by LLMs for automated claim verification with LLMs. While LLMs have demonstrated robust reasoning capabilities and human-readable explanations, they also pose significant threats through the generated misinformation, raising concerns about the trustworthiness of applying LLMs in claim verification tasks. \citet{huang2024position} also emphasized that transparency in both the models and the underlying technologies is crucial for fostering trustworthiness and proposed principles spanning multiple dimensions to examine LLMs' trustworthiness. Through an extensive analysis of 16 LLMs across over 30 datasets based on these principles, they found that proprietary LLMs generally outperform open-source models in trustworthiness, largely due to their superior functional capabilities.

\section{Conclusion}
In this work, we present an explainable biomedical claim verification system that integrates iterative human-AI collaboration, evidence-based rationales, SHAP saliency explanations, and comparative insights from multiple LLMs. Our approach introduces the \textbf{CoENLI} framework to improve transparency, accountability and adaptability so that domain experts can better trust and use the results provided by LLMs. SHAP values further clarify how specific components of the model-generated rationales contribute to the final prediction, enhancing interpretability and user confidence. We explore  the combination of advanced reasoning capabilities with the most advanced LLMs, such as GPT4o-mini and the open-source lightweight LLMs for interpretability, to achieve a balance between accuracy and transparency in the verification process. Additionally, our user study demonstrates the system’s practical benefits, as indicated by an increase in inter-annotator agreement and feedback emphasizing the usability and trustworthiness of the model's reasoning process. Nevertheless, there are areas for refinement of the intermediate reasoning steps to address user concerns about overlooked details in claims.

In future work, we aim to integrate this explainable verification system as a component within a broader evidence synthesis framework to support human-AI collaboration in tasks such as combating misinformation in healthcare domain.
Future work will also focus on refining intermediate reasoning quality, optimizing lightweight LLMs through task-specific fine-tuning to further enhance system performance, accessibility, and trustworthiness in biomedical claim verification and beyond.

\section*{Limitations} First, the reliance on GPT4o-mini in the \textit{Evidence-Based Evaluation} step imposes computational resource demands that may limit accessibility in low-resource settings. Furthermore, the current framework may still struggle with claims requiring highly nuanced or specialized domain knowledge, where additional fine-tuning or inclusion of external expert input may be necessary. Small-scale LLMs, though flexible, exhibit performance limitations without additional task-specific training data. Finally, while SHAP values provide an interpretive layer, their effectiveness depends on the quality and granularity of the generated rationales.

\section*{Acknowledgments}
This work was funded by the German Federal Ministry of Education and Research (BMBF) under grant numbers 01IW23002 (No-IDLE) and 01IW24006 (NoIDLEChatGPT), as well as by the Endowed Chair of Applied AI at the University of Oldenburg. We also appreciate the support by a grant from Accenture Labs. We also gratefully acknowledge the support provided by a grant from Accenture Labs\footnote{\url{https://iml.dfki.de/news/autoprompt-aims-to-improve-chatgpts-analysis-of-clinical-data/}}.
\section*{Declaration on Generative AI}
 During the preparation of this work, the authors used ChatGPT\footnote{\url{https://chatgpt.com/}} and DeepL\footnote{\url{https://www.deepl.com/en/write}} in order to: Grammar and spelling check. After using these services, the authors reviewed and edited the content and take full responsibility for the publication’s content.

\newpage

\newpage
\appendix
\section{Models}
\label{app:models}
\begin{table}[h!]
    \centering
  
    \begin{tabular}{llll}
    \toprule
       Model  & Version  & Context Window & Parameters  \\
       \midrule
       GPT3.5 &	gpt-3.5-turbo-0125 &	16K&	175B \\
        GPT4o-mini &	gpt-4o-mini-2024-07-18&	128K&	?\\
        \midrule
      
       Llama3.1-8B&Meta-Llama-3.1-8B-Instruct&	128K&	8B\\
       Gemma2-9B&	gemma-2-9b-bnb-it&	8K&	9B\\
Mistral-12B&	Mistral-Nemo-Instruct-2407&	1024K&	12B\\
Phi3-14B&	Phi-3-medium-4k-instruct &	4K&	14B\\
      \bottomrule
    \end{tabular}
    \caption{List of low-cost GPT models and lightweight open-source LLMs used in our experiments, and a comparison of model size and initial context window length. The model size of the open source LLMs is limited to 14 billion parameters. All models are the instruct fine-tuned version.}
    \label{tab:models}
\end{table}




\end{document}